\documentstyle[aps,preprint]{revtex}

\begin{document}
\draft


\title{Exclusion statistics for fractional quantum Hall states on a 
sphere}

\author{S. B. Isakov,$^{1,2}$, G. S. Canright,$^{1,3}$ 
and M. D. Johnson$^{4}$}
\vspace{1 ex}

\address{$^1$Senter for H\o yere Studier, Drammensveien 78, 
0271 Oslo, Norway \\
$^2$Department of Physics,  University of Oslo,
P.O. Box 1048 Blindern, 0316 Oslo, Norway\\
$^3$Department of Physics, University of Tennessee, 
Knoxville, TN 37996-1200 \\
$^4$Department of Physics, University of Central Florida, 
Orlando, FL 32816-2385}

\date{November 12, 1996}
\maketitle
\narrowtext
\tighten

\begin{abstract}
We discuss exclusion statistics parameters for quasiholes and 
quasielectrons  excited above the 
fractional quantum Hall states near    $\nu=p/(2np+1)$.
We derive the diagonal statistics parameters from the 
(``unprojected'') composite fermion (CF) picture. 
We propose values for  the off-diagonal (mutual) statistics 
parameters as a simple modification of those obtained from the 
unprojected CF picture,    
by analyzing  finite system numerical spectra in the spherical 
geometry.   

 \end{abstract}


\narrowtext

Excitations with a fractional charge --- quasiholes 
(QHs) and quasielectrons (QEs) --- arising in the fractional quantum
Hall  (FQH) effect \cite{book-QHE} have been shown to obey     
{\em fractional exchange} statistics, i.e. they are anyons 
\cite{anyons}.
The latter concept is two-dimensional.
However the system in the FQH effect is restricted to the 
lowest Landau level (LL),  
which makes it effectively one-dimensional. 
In this case an algebraic definition of {\em fractional statistics 
in one dimension}\cite{LM-Heis} applies to anyons \cite{HLM}.  
One more definition  of fractional statistics, {\em
exclusion statistics\/} (ES),  
which explicitly specifies  the dimension of the
many-particle Hilbert space, was introduced and applied to QHs 
and QEs \cite{Haldane-ES}. 
Connection between the two last definitions is established in 
statistical mechanics:
the statistical distribution for 1D fractional 
statistics~\cite{I-IJMPA}
coincides with that for exclusion statistics\cite{I-MPLB,Wu}.
The ES counting directly applies to anyons in the lowest 
LL \cite{CJ}. Relation of the thermodynamics of anyons in the lowest 
LL~\cite{dVO1} to that of ES  was also shown \cite{Wu} 
including mutual statistical interactions \cite{SSS}.

The ES counting of states implies 
for the many-particle Hilbert space dimension
($\alpha =+$ and $-$ refer to QEs and QHs, respectively) 
\cite{Haldane-ES}
\begin{equation}
W=\prod_{\alpha}{D_{\alpha} +N_{\alpha} -1\choose N_{\alpha}},
\label{W-ES}\end{equation}
where 
$D_{\alpha}=d_{\alpha}-\sum_{\beta}g_{\alpha\beta}
(N_{\beta}-\delta_{\alpha\beta})$, 
and $d_{\alpha}$ are   
the dimensions of the one-particle Hilbert spaces \cite{CJ,JC}.

Based on  heuristic duality arguments, Haldane\cite{Haldane-ES} 
originally proposed the following values for the  
statistics parameters:
$g_{--}=g_{+-}=-g_{++}=-g_{-+}=1/m$ for 
Laughlin $\nu=1/m$ states ($m$ odd).
However, analysis of numerical spectra for a small number of 
electrons on a sphere for states near $\nu=1/3$ led to the 
conclusion \cite{JC} that 
$g_{--}=\frac13$ but  $g_{++}=\frac53$. 
The origin of the  asymmetry  
between $g_{--}$ and $g_{++}$  
was attributed to the interpretation of 
QEs as hard core anyons, resulting in  
$g_{--}={1}/{m}$, $g_{++}=2-{1}/{m}$ for Laughlin states
\cite{HXZ,LiO,Yang}.
Very recently, new values also for the off-diagonal statistics 
parameters,  
$g_{-+}=-g_{+-}=2-{1}/{m}$, have been 
proposed \cite{SWY}. 


In this paper we discuss the  ES  parameters 
for quasiparticles for the more 
general filling factors $\nu= p/(2np+ 1)$, with $n$ and $p$ positive 
integers,
exploiting the notion of composite fermions (CFs) \cite{CFs}. 
We discuss   QHs and QEs at the highest
($p$-th)   level of the hierarchy if  the $p/(2np+ 1)$ states  
are viewed as hierarchical states. In the CF language it means that 
in addition to $p$ completely filled CF LLs there are CFs in the 
$(p+1)$-th CF 
LL (QEs) and empty states in the $p$-th CF LL (QHs).

Exact diagonalization data  for FQH states 
for a few  electrons on a sphere\cite{HaldanePRL83,Haldane-QHE} 
demonstrate 
a band structure of the energy levels 
\cite{HXZ,DevJain92,WuJainPRB95}.
We use the  observation -- made for $n=p=1$ (near the  $\frac13$ 
state) -- that  the lowest band of levels is recovered correctly 
by the (``unprojected'') CF picture\cite{DevJain92} 
to derive the diagonal ES parameters. 
We then verify these parameters numerically  
for larger values of $n$ and $p$.  
We also propose off-diagonal 
statistics parameters for which the ES counting (\ref{W-ES})  
recovers correctly the number of states in first excited bands 
corresponding to the above configurations of QHs and QEs,
for all the  numerical data available. 


Consider CFs, each carrying   $2n$ flux quanta,  on a sphere. 
The total magnetic flux through the sphere affecting a CF is an 
integer number $2S$ of quanta from the monopole charge minus 
$2n (N_e-1)$ flux quanta bound to other CFs:
\begin{eqnarray}
2 {S^*}= 2S - 2n (N_e-1)\;.
\label{DeltaS}\end{eqnarray}   
``Unprojected'' CFs occupy the energy levels in the effective 
magnetic field determined by $2S^*$. 
The relevant eigenstates are given by the monopole spherical 
harmonics  $Y_{lM}^{({S^*})}$, 
where $l={S^*},{S^*} +1, \ldots $ numbers 
the ``Landau levels'' on the sphere, 
and $M$ taking values $-l,\ldots,l$ counts the degeneracies of 
these levels \cite{WuYang}.
When $p$ CF LLs are filled, 
$N_e=2p{S^*}+p^2$, which results in the filling factor  
$\nu={N_e}/{2S}\to {p}/({2np+1})$ as  $S\to \infty$.

If there are $N_-$ QHs and $N_+$ QEs,  
\begin{eqnarray}
N_e=2p{S^*} +p^2 -N_{-}+N_{+}\;.
\label{flux-balancing}\end{eqnarray}  
Here a quasihole 
is an empty state in the $p$-th CF LL (with angular
momentum  $l_{-}={S^*} +p-1$), 
a quasielectron is a CF in the $(p+1)$-th CF 
LL (with angular momentum $l_{+}={S^*} +p$) \cite{QPs}. 
The dimension of the  Hilbert space for $N_-$ QHs and $N_+$ QEs  
thus factors as
\begin{eqnarray}
W'={2 l_{-}+1 \choose N_-}{2 l_{+}+1\choose N_+} \;.
\label{W'}\end{eqnarray} 

Comparing  this (with ${S^*}$ determined by (\ref{DeltaS}) and 
(\ref{flux-balancing})) with the 
ES counting of states (\ref{W-ES}) yields 
the dimensions of one-QH and one-QE states   
\begin{equation}
d_-=\frac{2S+p+4n}{2np+1}+p-1 , \ \ 
d_+=\frac{2S+p}{2np+1}+p+1 
\label{d1}\end{equation}  
as well as  the statistics matrix
\begin{eqnarray}
g'_{++}&=&2-g'_{--}=1+{2n}/({2np+1})\;,
\label{G'-diag}\\
g'_{-+}&=&-g'_{+-}={2n}/({2np+1})\;.  
\label{G'-off-diag}\end{eqnarray}

The lowest band of levels corresponds to the case where  
$|N_+ - N_-|$ in (\ref{flux-balancing}) takes its minimal possible 
value (so that  at least one of the numbers 
$N_-$ and $N_+$ vanishes). Then (\ref{W-ES})   implies that the 
 number of states in the  band 
is completely determined by the diagonal ES parameters.

It has been demonstrated numerically near  $\nu=\frac13$  
($n=p=1$) that the unprojected CF counting (\ref{W'}) 
(and correspondingly, the ES counting 
with the parameters (\ref{G'-diag})) recovers correctly the  number
of states in the lowest band of levels \cite{DevJain92}.
We have also verified the parameters (\ref{G'-diag}) for larger 
values of $p$.   Examples are  the spectra in Fig.~1. In Fig.~1(a),  
the lowest band of 21 states corresponds to 
2 QHs near $\nu=\frac13$.
Fig.~1(b) represents the spectrum which gives support for the 
diagonal 
statistics parameters (\ref{G'-diag})  for larger values of~$p$. 
The lowest band 
consists of a single multiplet of angular momentum $\frac52$. The 6 
states are recovered by the ES counting
if they are   viewed either  as corresponding to 
{\em one} QE near $\nu=\frac25$ ($n=1$, $p=2$), 
or as corresponding to {\em five} QHs   
with respect to the state $\nu=\frac37$ ($n=1$, $p=3$).    
For numerical support of  the parameters (\ref{G'-diag}) for larger 
values of $n$, see discussion of Fig.~2 below for  $\nu=\frac15$.

Based on the above, we expect that  the diagonal ES parameters 
derived from
the unprojected CF picture (\ref{G'-diag})
are generally valid, that is 
$g_{++}=g'_{++}$ and $g_{--}=g'_{--}$.   
For Laughlin states ($p=1, m=2n+1$), 
(\ref{G'-diag}) reduce to those obtained in 
Refs.~\onlinecite{JC,LiO,Yang}.

Let us now turn to excited bands that are obtained from the lowest 
band by adding a pair (1 QH +1 QE).   
The {\it full\/} CF picture 
is obtained from the unprojected CF picture by  
the projection onto the lowest LL of the real magnetic field; 
the latter procedure  eliminates some of the states counted in 
(\ref{W'}), 
as was shown numerically in various cases
\cite{DevJain92,WuJainPRB95}.
To account for this, we modify the state counting (\ref{W'}), 
introducing additional ``statistical interactions'' between QHs and 
QEs, supposing  that excitations of one kind affect the number of 
states available for excitations of the other kind as follows:

\begin{eqnarray}
W&=&{2 l_{-}+1-\tilde g_{-+}N_{+}\choose N_-}
{2 l_{+}+1-\tilde g_{+-}N_{-}\choose N_+} \;.
\label{W'modified}\end{eqnarray} 
This modifies the off-diagonal elements of the statistics matrix to  
\begin{eqnarray}
g_{-+}=g'_{-+} +\tilde g_{-+}\;,\quad 
g_{+-}=g'_{-+} +\tilde g_{+-}\;.
\label{G-offdiag}\end{eqnarray}


Consider first the case with 
$N_-=N_+=1$ 
in the first  excited band (so that the lowest band is precisely the 
$p/(2np+1)$ state with zero angular momentum).
According to  the unprojected  CF picture, 
angular momenta of a QE and a QH ($l_+$ and $l_-$) differ by unity. 
Hence the angular momentum of the  pair 
(coinciding with the total angular momentum of the state) 
may take values $L=1,2,\ldots, L_{\rm max}$, where 
$L_{\rm max}=l_{+}+l_{-}=N_e/p+p-1$.

Relevant numerical data are collected in Table 1. 
The first neutral excited band always contains a single multiplet 
for each 
angular momentum $L=2,3,\ldots , L_{\rm max}$. 
As seen from Table 1, the 
maximal angular momentum in the band 
is recovered correctly in the unprojected CF picture.
The difference from the latter picture is in  
the {\it minimal\/} value of the total angular
momentum, which is {\em two} (rather than {\em one}) for 
the system of interacting electrons.
One can provide general  arguments in support of the latter  
observation \cite{He};
note also that the projection onto the lowest LL 
in the CF picture 
eliminates  the state with $L=1$, as was shown numerically for
$\nu=\frac13$ \cite{DevJain92}.

One can recover the correct number of states in the first neutral 
excited 
band, including  the decomposition of states in  
angular momentum, within the ES counting. To show this, 
we formally introduce off-diagonal ES parameters depending 
on an integer number $\ell$, keeping their  antisymmetry,
\begin{equation}
g_{-+}(\ell )=g'_{-+} + \ell , \ \  g_{+-}(\ell )=g'_{+-} - \ell ,   
\label{gofl}\end{equation}
and consider the associated ``partial'' statistical weights  
[cf.\ (\ref{W-ES})]
corresponding to these statistics parameters:
\begin{eqnarray}
W(\ell )=[d_{-}-g_{-+}(\ell )][d_{+}-g_{+-}(\ell)]\;.
\label{W(l)}\end{eqnarray} 
It then follows from (\ref{d1})--(\ref{G'-off-diag}) that 
$W(\ell)-W(\ell+1)= 2(\ell +1)+1$, and  one can  write  
\begin{eqnarray}
W(\ell)=\sum_{L=\ell +1}^{L_{\rm max}} (2 L +1) \;.
\label{W(l)=}\end{eqnarray}
If one identifies $L$ with the angular momentum of the pair (equal to
the total angular momentum), then formula (\ref{W(l)=}) manifests the
angular momentum decomposition in the ES counting: the
difference $W(\ell)-W(\ell+1)$ counts states 
with angular momentum $L=\ell +1$~\cite{nesting}.

We observe  that  $W(1)$ recovers the correct 
number of states in the lowest excited band. The corresponding 
off-diagonal statistics parameters are hence given by   (\ref{gofl})
with $\ell =1$. 
Comparing this with (\ref{G-offdiag}) then yields 
$\tilde g_{-+}=-\tilde g_{+-}=1$,
thus finally resulting in  the ES matrix 

\begin{eqnarray}
g_{++}=g_{-+}=-g_{+-}=2-g_{--}=1+ \frac{2n}{2np+1}\;.
\label{G}\end{eqnarray}  
For Laughlin states,  the off-diagonal statistics parameters 
in  (\ref{G}) reduce to those proposed in 
Ref.~\onlinecite{SWY}.

The above logic entirely determines the set of statistics parameters
$g_{\alpha\beta}$, using as input only the bands with a single
species (thus determining the ``diagonal'' elements), and the
first neutral excited bands with $N_+ = N_- = 1$. 
It is then of course of interest 
to try to apply these parameters to the other excited bands.
Here we have to demand that the band should correspond to a  
a unique configuration (the numbers $(N_-, N_+)$)
of quasiparticles   
  (for the case involving several 
configurations, see discussion of Fig.~3 below). 
The above first  neutral excited band, which we 
refer to as case (i), satisfies this 
requirement. Another case, which is referred to as case (ii), is the 
first excited band containing one QE and any number of QHs 
around the $1/m$ state.      

We have examined a large number of spectra for case~(ii) 
around the $\frac13$ state  (some of which may be 
found in previously published spectra
\cite{HXZ,WuJainPRB95,Quinn,SWY}. 
Since case (ii) fixes $N_+ = 1$,      
we will simply list the cases examined as ($N_e,N_-$) (with $N_-$ the
number of QH in the {\it first excited\/} band):
($N_e=5$, $N_-=1,2,3,4$); ($N_e=6$, $N_-=1,2,3,4$); ($N_e=7$,
$N_-=1,2,3$).
One of these bands ($N_e=5$, $N_-=3$) is clearly
seen in Fig.~1(a). In this, and in 
{\it all other\/} 
of these cases, the ES counting (\ref{W-ES}), as specified by 
(\ref{d1}) and (\ref{G}),
correctly counts the states in the band, which fall below a
well-defined (though small in a few cases) gap in the spectrum.

For FQH states with $n>1$, a band structure is 
expressed in the numerical data on a smaller energy scale.
Thus, viewing the spectrum in Fig.~2 in terms of excitations around 
the $\frac15$ state yields $N_{-}=1$ in 
the lowest band, and $N_{-}=2$, $N_{+}=1$ in the first excited band. 
The ES counting, using (\ref{G}),
then yields 6 states in the lowest band and 100 states in the first 
excited band. These two sets of
states may be identified in Fig.~2 \cite{coarse}.

As for case (i), the ES counting for case (ii) is in agreement with 
the full CF picture: it reproduces 
165 states in the first excited band,
obtained numerically in Ref.~\onlinecite{WuJainPRB95}
(after   the projection onto the lowest LL) 
for $N_e=6$, $2S=16$ 
($N_{-}=2$, $N_{+}=1$ around $\nu=\frac13$  in the first 
excited band). 
The simple general formulas for the number of states 
in bands (ii)  (e.g. in the form (\ref{W'modified})) thus 
call for an analytical understanding of 
the CF projection procedure, including  underlying symmetry 
properties of the CF wave functions.  

For cases other than (i) and (ii),
the counting of states in the low-lying excited bands is more 
complicated. Consider, e.g, the case presented in Fig.~3.
In the CF picture,  the (degenerate) ground state has then the 
lowest LL filled, and one CF in the
second LL. Therefore there are two distinct ways to form states in 
the first excited band 
(and consequently, two distinct {\em configurations} of 
quasiparticles relevant  to this band), 
 by exciting a CF (a) from the first to the second LL, or
(b) from the second to the third LL.
The total number of states in the band 
is expected to be the sum of two terms.

The ES counting, with (\ref{G}) and (\ref{d1}), applies to (a), 
yielding  
84 states ($N_+=2$, $N_-=1$), whereas the total number of states in 
the first excited band in Fig.~3  is 93. 
The difference (9~states) then represents a term of type (b), 
and this is precisely 
the Hilbert space dimension for a single CF in the third
LL \cite{Wu-counting}. 
The latter remark also shows the way to study other excited bands 
and, correspondingly, evaluate the density of excited states in 
energy for FQH states, using the ES counting. Work 
along these lines is in progress.

In conclusion, we have
found the exclusion statistics parameters, both diagonal and 
off-diagonal, for QHs and QEs 
at the highest level of the hierarchy, for FQH states near 
$\nu=p/(2np+1)$. With these parameters, the ES counting of
states applies to the first neutral excited band for the 
above FQH states on a sphere,
as well as to the first excited bands containing one
QE and any number of QHs around the $1/m$ state. 
This  provides  general formulas for the number of states in the 
above bands, which are in agreement 
with numerical results given by the full (``projected'') CF picture. 

We thank Daniel Arovas, Heidi Kj\o nsberg, Jon Magne Leinaas, 
Stefan Mashkevich, Jan Myrheim, John Quinn, and Yong-Shi Wu
for helpful discussions, and Piotr Sitko for
sharing some unpublished results.
SBI and GSC are grateful to the Senter for H\o yere Studier
 in Oslo, where this work was completed, 
for warm hospitality and support.
GSC was also supported in part by the NSF under grant DMR-9413057,
and MDJ by the NSF under grant DMR-9301433.

%
%




\begin{figure}
\caption{Low-lying energy levels for electrons on a sphere: 
(a) $N_e = 5$, $2S=14$.
The solid lines
separate the lowest  band corresponding to two  
QHs (near $\nu=\frac13$)
from the first excited band (3~QHs + 1~QE) and the latter band from
higher states.
(b) $N_e = 7$, $2S=13$.
The lowest  ``band'' is a single 6-plet, whose dimension
is obtained using diagonal ES parameters (6),
with $p>1$.}
\end{figure}

\begin{figure}
\caption{Low-lying states for $N_e = 5$, $2S=21$ (one QH near 
$\nu =\frac15$ 
in the lowest band).
The two lowest bands are counted
correctly by the ES counting, with $n=2$, $p=1$.}
\end{figure}

\begin{figure}
\caption{Low-lying states for $N_e = 6$, $2S=14$: 
the first excited band corresponds to two different configurations of 
quasiparticles (see text). }
\end{figure}


\begin{table}
\caption{Neutral (1QE+1QH) bands above states
$\nu=p/(2np+1)$, for various $N_e$ and $\nu$.} 
\begin{tabular}{ccccc}
$N_e$  & $2S$ & $\nu$ &$L_{\rm max}$ &  Ref. \\
\tableline
6    & 15 & 1/3 & 6          & \onlinecite{KWJ} \\
7    & 18 & 1/3 & 7          & \onlinecite{KWJ} \\
8    & 21 & 1/3 & 8          & \onlinecite{HXZ} \\
9    & 24 & 1/3 & 9          & \onlinecite{Fano} \\
5    & 20 & 1/5 & 5          & \onlinecite{Haldane-QHE}\\
5    & 28 & 1/7 & 5          & \onlinecite{Haldane-QHE}\\
8    & 16 & 2/5 & 5          & \onlinecite{KWJ} \\
10   & 21 & 2/5 & 6          & \onlinecite{Morf} \\
9    & 16 & 3/7 & 5          & \onlinecite{KWJ} \\
12   & 23 & 3/7 & 6          & \onlinecite{He} \\
\end{tabular}
\end{table}


\begin{references}
\bibitem{book-QHE}  
{\em The Quantum Hall Effect}, ed. by R. E. Prange and 
S. M. Girvin (Second Edition, Springer, 1990).

\bibitem{anyons} {\em Fractional Statistics and
     Anyon Superconductivity}, ed. by F. Wilczek, World
     Scientific (1990).

\bibitem{LM-Heis} J. M. Leinaas and J. Myrheim, Phys. Rev.~B 
{\bf 37}, 9286 (1988); Int. J. Mod. Phys. B {\bf 5}, 2573 (1991).

\bibitem{HLM} T. H. Hansson, J. M. Leinaas, and J. Myrheim, 
Nucl. Phys. B {\bf 384}, 559 (1992).

\bibitem{Haldane-ES} F. D. M. Haldane, Phys. Rev. Lett. {\bf 67}, 937
(1991).

\bibitem{I-IJMPA} S. B. Isakov, Int. J. Mod. Phys. A {\bf 9}, 2563
(1994).

\bibitem{I-MPLB} S. B. Isakov, Mod. Phys. Lett. B {\bf 8}, 319 
(1994).

\bibitem{Wu} Y. S. Wu, Phys. Rev. Lett. {\bf 73}, 922 (1994).

\bibitem{CJ} G. S. Canright and M. D. Johnson, J. Phys. A {\bf 27},
3579 (1994).

\bibitem{dVO1} A. Dasni{\`e}res de Veigy and S. Ouvry,
Phys. Rev. Lett. {\bf 72}, 600 (1994).

\bibitem{SSS} S. B. Isakov, S. Mashkevich, and  S. Ouvry,
Nucl. Phys. B {\bf 448}, 457 (1995).  

\bibitem{JC} M. D. Johnson and G. S. Canright, Phys. Rev. B {\bf 49},
2947 (1994).

\bibitem{HXZ} S. He, X. C. Xie, and F. C. Zhang, Phys. Rev.
Lett. {\bf 68}, 3460 (1992).

\bibitem{LiO} D. Li and S. Ouvry, Nucl. Phys. B {\bf 430}, 563 
(1994).

\bibitem{Yang} J. Yang, Phys. Rev. B {\bf 50}, 11 196 (1994). 

\bibitem{CFs} J. K. Jain, Phys. Rev. Lett. {\bf 63}, 199 (1989);
Phys. Rev. B {\bf 41}, 7653 (1990); Science {\bf 266}, 1199 (1994).

\bibitem{SWY} W.~P.~Su, Y.~S.~Wu, and J.~Yang,  
Phys. Rev. Lett. {\bf 77}, 3423 (1996).

\bibitem{HaldanePRL83} F. D. M. Haldane, Phys. Rev. Lett. {\bf 51}, 
605 (1983).

\bibitem{Haldane-QHE} F. D. M. Haldane, Chapter 8 of Ref.~1. 

\bibitem{DevJain92} G. Dev and J. K. Jain, Phys. Rev. Lett. {\bf 69},
2843 (1992). 

\bibitem{WuJainPRB95} X. G. Wu and J. K. Jain, Phys. Rev. B {\bf 51},
1752 (1995).

\bibitem{WuYang} T. T. Wu and C. N. Yang, Nucl. Phys. B {\bf 107},
365 (1976); Phys. Rev. D {\bf 16}, 1078 (1977).

\bibitem{QPs} If Jain's states with $\nu=p/(2np+1)$ are viewed as 
hierarchical 
states, this definition of QPs is  apparently in agreement with the 
definition of QPs at the $p$-th level of the hierarchy.  

\bibitem{KWJ} R. K. Kamilla, X. G. Wu, and J. K. Jain, 
Phys. Rev. Lett.
{\bf 76}, 1332 (1996); Report cond-mat/9604023.

\bibitem{Fano} G. Fano, F. Ortolani, and E. Colombo, Phys. 
Rev. B {\bf 34}, 2670 (1986).

\bibitem{Morf} N. d'Ambrumenil and R. Morf, Phys. Rev. B {\bf 40},
6108 (1989).

\bibitem {He} S. He, S. H. Simon and B.I. Halperin, Phys. Rev. B 
{\bf 50}, 1823 (1994).

\bibitem{nesting} For $N_+>1$ and/or $N_->1$, one can show the more
general ``nesting'' property that $W(\ell'>\ell) < W(\ell)$.

\bibitem{Quinn}  P.~Sitko, S.~N.~Yi, K.~S.~Yi, and J.~J.~Quinn, 
Phys. Rev. Lett. {\bf 76}, 3396 (1996). 

\bibitem{coarse}
Fig.~2 may also be interpreted as 
excitations around $\nu=\frac13$, for which one expects 9 QHs and so
2002 states [using (\ref{W-ES})]. 
These states are the entire lowest parabolic band in
Fig.~2. Thus we see, in this figure, both a `coarse' and a `fine'
structure in the energy bands, with the corresponding energy scales
determined by the Coulomb pseudopotentials\cite{Haldane-QHE}
$V_1$ and $V_3$, respectively.

\bibitem{Wu-counting} 
We thank Y.~S.~Wu who pointed out the latter  observation to us.  


\end{references}
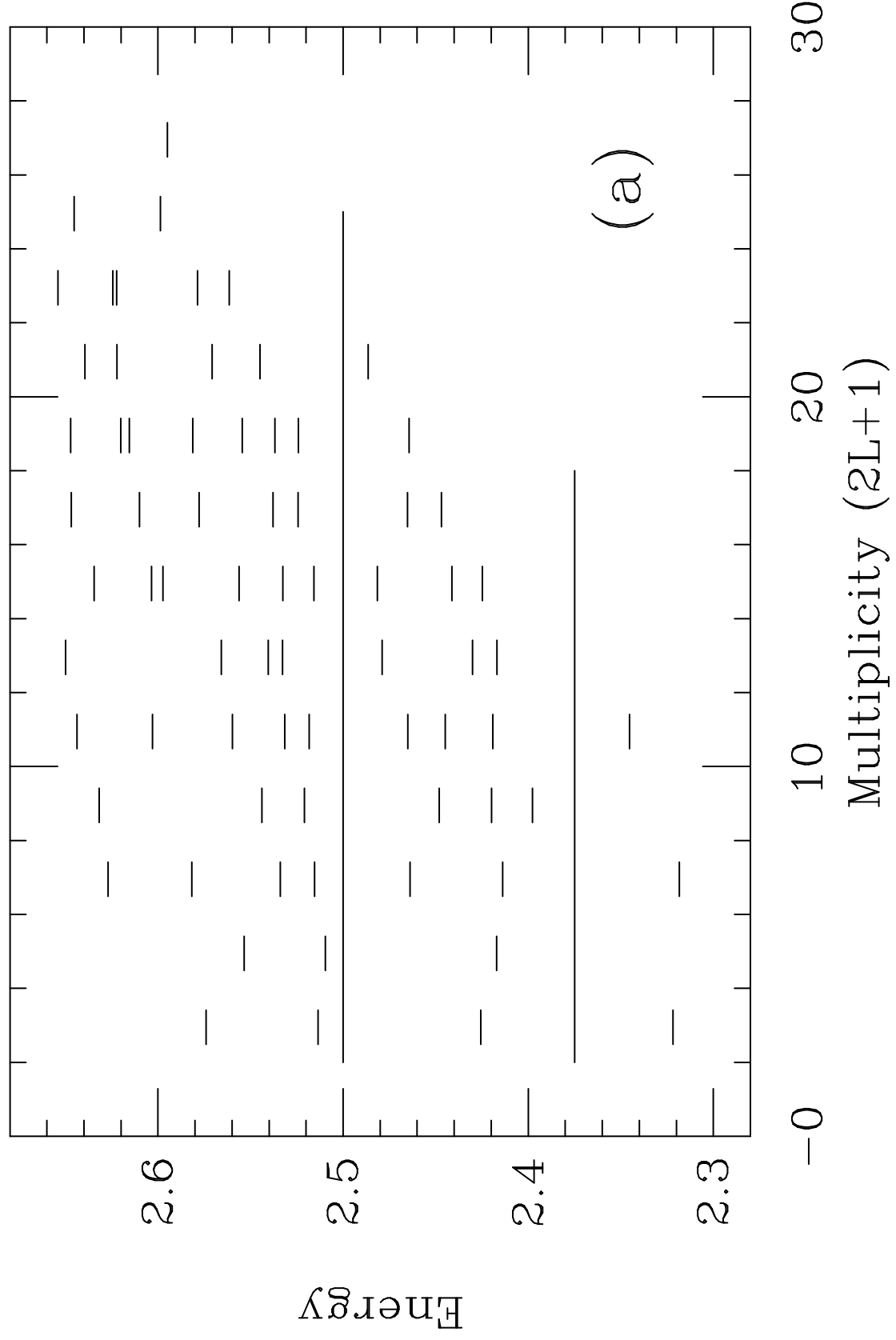
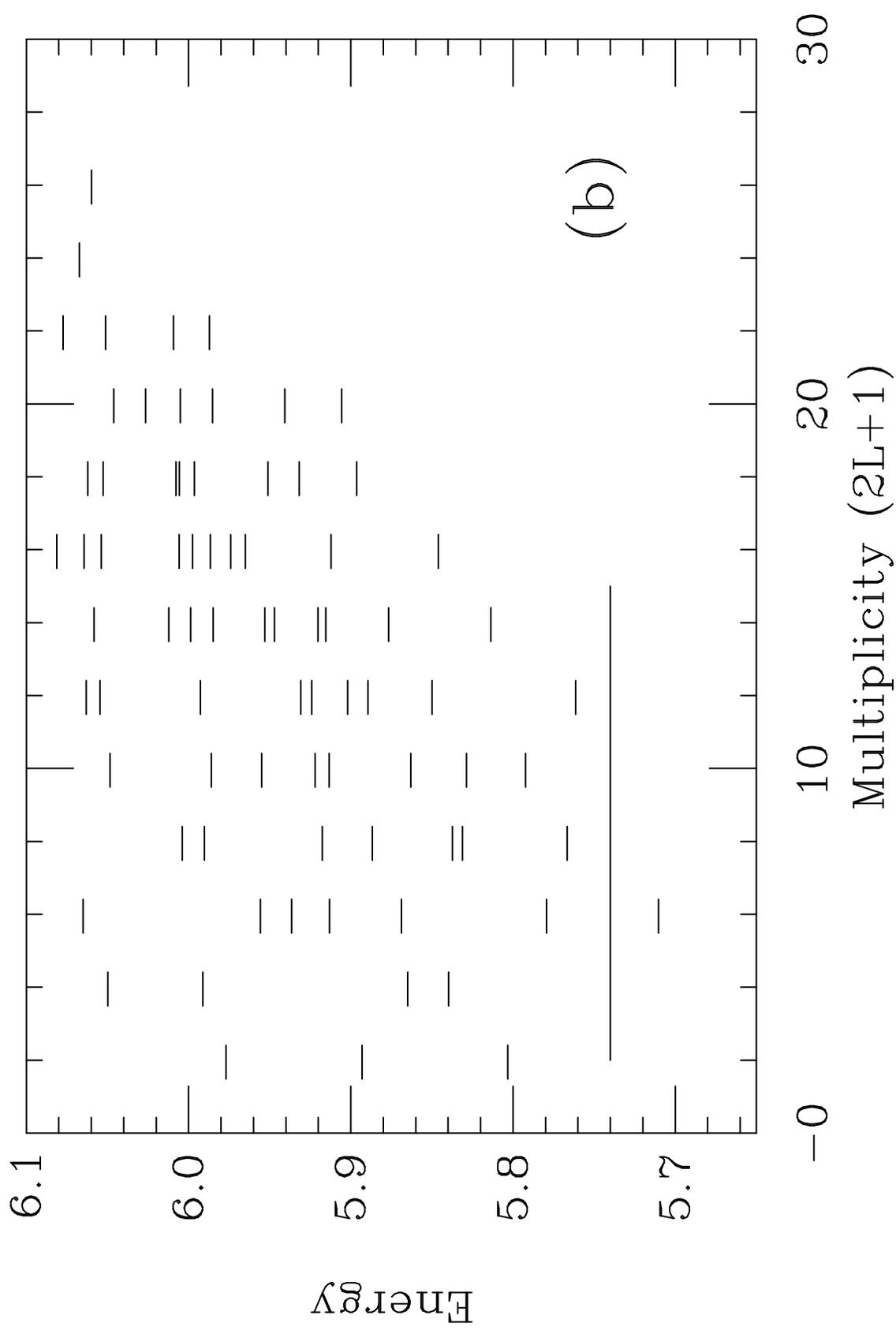
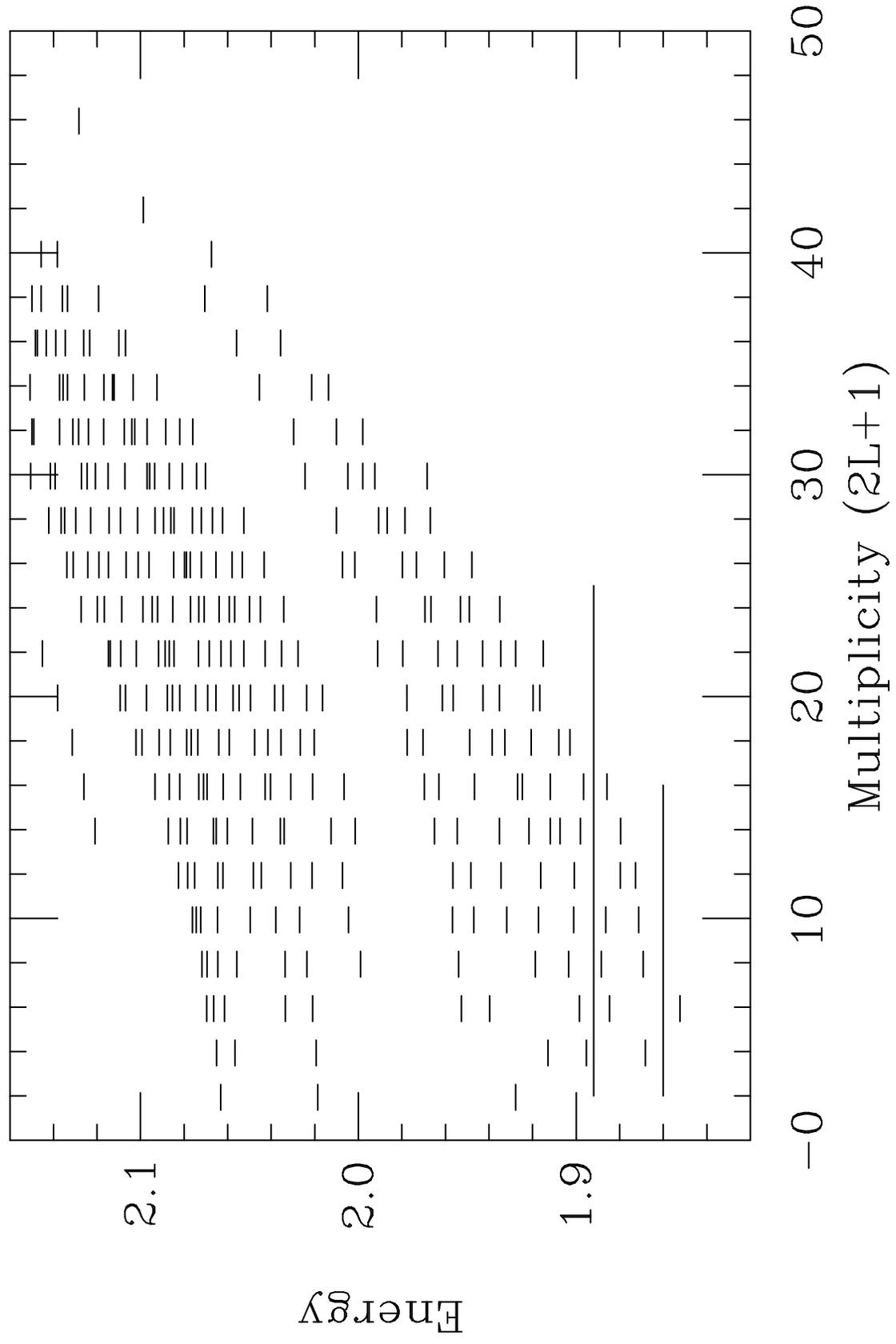
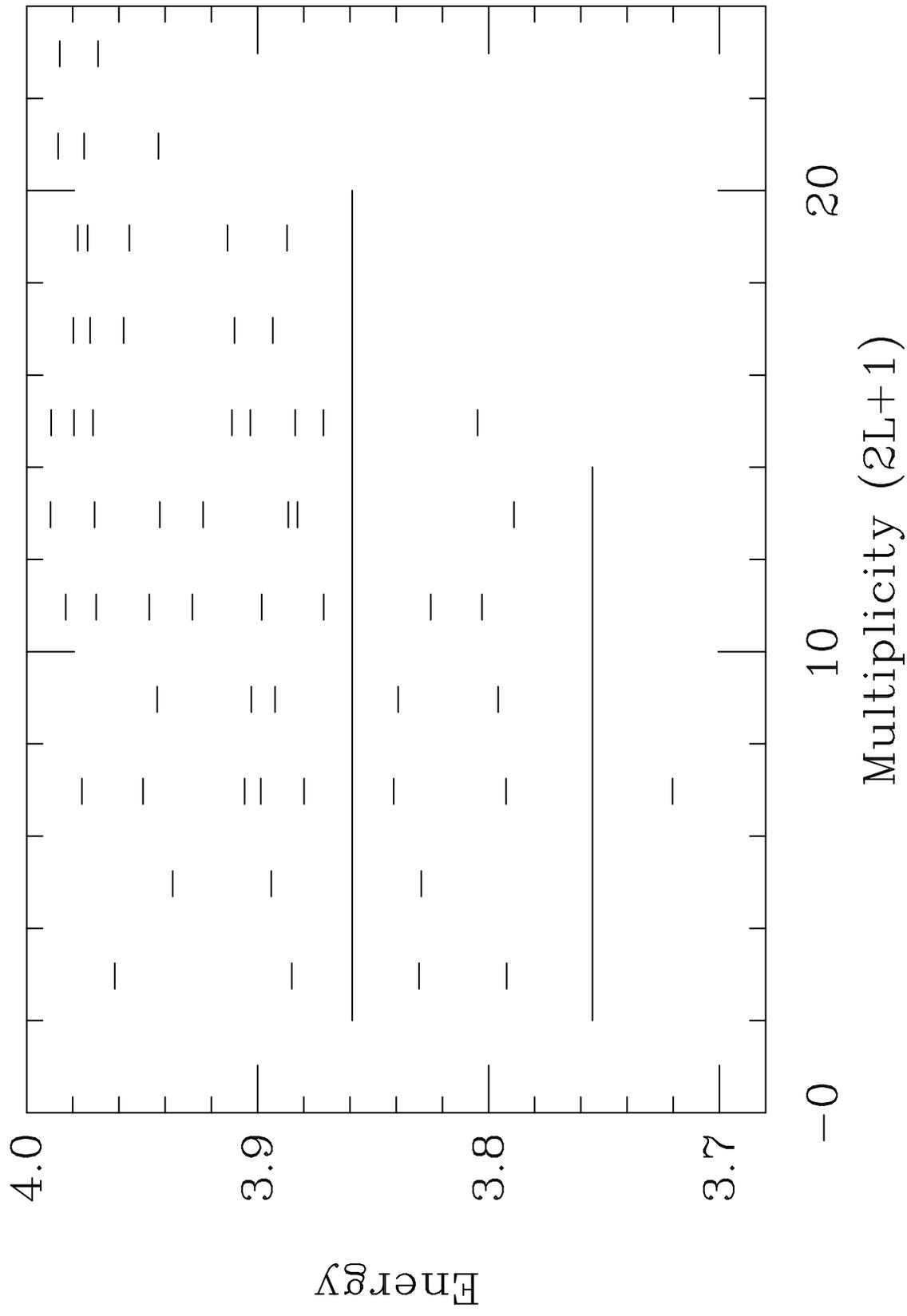
\end{document}